\documentclass{llncs}

\usepackage{amsmath}
\usepackage{amsfonts}

\usepackage{graphicx}
\usepackage{color}

\usepackage[all]{xy}

\def\ligne#1{\hbox to \hsize{\hfill#1\hfill}}

\def\ie{\emph{i.e.}}

\def\loopc#1{\ensuremath{\nabla #1}}

\newcommand{\di}{\diamond}

\newcommand{\si}[1]{\ensuremath{\mathcal #1}}

\newcommand{\op}{\textstyle\bigoplus}

\def\complex#1#2{\hbox{\ensuremath{#1\otimes #2}}}

\newcommand{\To}{~~\hbox{\Large$\vdash$}~~}

\title{DNA Circuits Based on Isothermal Constrained Loop Extension DNA Amplification}

\author{Maurice Margenstern\inst{1}\and Pascal Mayer\inst{2}\and
Sergey Verlan\inst{3}}

\institute{
Universit\'e Paul Verlaine - Metz, LITA, EA 3097, UFR MIM,\\
Ile du Saulcy, 57045 Metz C\'edex, France\\
and CNRS, LORIA,\\
E-mail: \email{margens@univ-metz.fr,mmargens@loria.fr}
\and
 BioFilm Control SAS,\\
 Biop\^ole Clermont Limagne,\\
 63360 Saint-Beauzire, France.\\
 Email: \email{pascal.mayer@neuf.fr}
\and
LACL, D\'epartement Informatique, Universit\'e Paris 12\\
61 av. G\'en\'eral de Gaulle, 94010 Cr\'eteil, France \\
E-mail: \email{verlan@univ-paris12.fr}
}

\begin{document}
\maketitle

\begin{abstract}
In this paper, we first describe the isothermal constrained loop extension DNA
amplification (ICLEDA), which is a new variant of amplification combining the
advantages of rolling circle amplification (RCA) and of strand displacement
amplification (SDA). Then, we formalize this process in terms of the theory of
formal languages and show, on the basis of this formulation, how to manage OR
and AND gates. We then explain how to introduce negation, which allows us to
prove that, in principle, it is possible to implement the computation of any
boolean function on DNA strands using ICLEDA.
\end{abstract}



\section{Introduction}

The first attempt to use DNA for solving computational problems was done by
Adleman~\cite{Adleman94}. 
Since that time several models of computation using DNA
have been proposed, we refer to~\cite{Amos05} for an overview. Boolean circuits
play an important role in this research. Their structure
allows us to implement
them in a simple way on DNA support.  There are several designs simulating
bounded fan-in circuits~\cite{OR99} and semi-unbounded fan-in
circuits~\cite{OR98} (in both cases AND and OR gates are used). Another
approach 
can be found in~\cite{AGH97} where circuits with NAND gates are
simulated or in~\cite{ZARP09} where the construction is based on the operation
of hybridization of molecular beacons.

In this article we use an approach similar to~\cite{OR98,AGN05}. More
precisely, we use only true values and the true value of any gate will be
encoded by a specific DNA sequence. However, we don't use hybridization to
assembly a resulting (answer) molecule like in these papers. Instead we use a
special type of amplification to express the presence of signals (true values
for some gates) which can further trigger the amplification of new signals
following the circuit. Such an approach does not require temperature cycles and
can be executed autonomously.

The most common in-vitro DNA strand replication method (also called $\!$``DNA
amplification'') is based on PCR. This method is based on a series of primer
extension cycles with changing temperature conditions to allow for strand
separation at the beginning of each cycle.

Rolling-circle amplification (RCA) is another method of strand replication
based on circular DNA molecules~\cite{FX95} and is inspired from the natural
replication mechanisms of some viruses~\cite{GD68}. The important observation
is that this method does not require changing the conditions of the test tube
for DNA amplification and produces long single stranded DNA molecules
including multiple complementary copies of the circular template DNA fragment.
This procedure was used in DNA computing as a basis for the simulation of the
resolution refutation in~\cite{LPCZ03}.

The strand displacement amplification (SDA) is based on the ability of a
restriction enzyme to nick a modified recognition site and the ability of a
polymerase to initiate synthesis at the nick and displace a downstream DNA
strand during replication~\cite{WLNS92}.
Both above methods allow us to produce
DNA strands in isothermal conditions. There are methods using both RCA and SDA,
for example ramification-extension method (RAM)~\cite{ZBHL98}.

In this article we consider a new isothermal DNA replication method, called
ICLEDA for "Isothermal Constrained Loop Extension DNA Amplification", described
in~\cite{brevet}. It 
makes it possible to produce short linear and single stranded DNA strands in
isothermal conditions. Importantly in the perspective of a practical
application, the amplification is also possible when the template molecules are
immobilized on a support.  We formalize the amplification process in terms of
formal languages. Such a formal system is constructed from a number of
elements, which we call \emph{amplification loop complexes} (or simply loop
complexes) that can be in
two states: blocked or unblocked. A loop complex in 
unblocked state produces
infinitely the corresponding DNA strand (signal). The transition from a blocked
to an unblocked state is done by annealing and primer extension. As a result we
can simulate a signaling cascade whose nodes correspond to AND and OR gates.
The result is collected in one of the two output nodes, corresponding to the
true or false value of the corresponding boolean function.

We also consider a more general framework concerning double-stranded DNA
molecules that are partially hybridized and that can be dissociated by
annealing with other single stranded DNA and/or by primer extension. We give a
description of the corresponding objects and operations in terms of the formal
language theory. This notation gives us a simple way to describe the simulation
of logical gates and the construction of the circuit.



\section{The mechanism}

   In this section, we first describe the ICLEDA amplification process
defined in \cite{brevet} on which the whole work is based.  Then, we show how
this mechanism allows us to devise a configuration, which we call the {\bf loop
complex}, which will later on allow us to implement logical gates in this
context. Note that the word loop refers to the shape of the biophysical complex
we consider rather than the computational device which is usually understood by
this term, this is why the term {\it complex} is attached to {\it loop} in this
denomination. The bio-physical description of the amplification process and of
the loop complex is the content of Subsection~\ref{ssec:ampli_loop}.

   In a second subsection, we propose a formalization of the process
described in Subsection~\ref{ssec:ampli_loop}, see Subsection~\ref{ssec:formal}.
Later, we shall switch to a bit more abstract formalism which will be more
suited for computation purposes.

\subsection{The amplification and the loop}
\label{ssec:ampli_loop}

The ICLEDA amplification method designed in patent~\cite{brevet} is to some
extend a combination of RCA and SDA amplification. We refer to this patent for
more technical information.

\begin{figure}[htb]
\begin{center}
\strut\hfill
\parbox{30mm}{
\begin{center}
\includegraphics[scale=0.5]{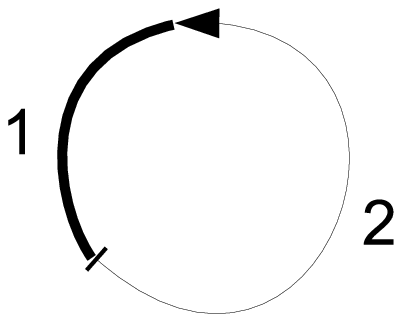}
\\[5mm]
(a)
\end{center}
}
\hfill
\parbox{30mm}{
\begin{center}
\includegraphics[scale=0.5]{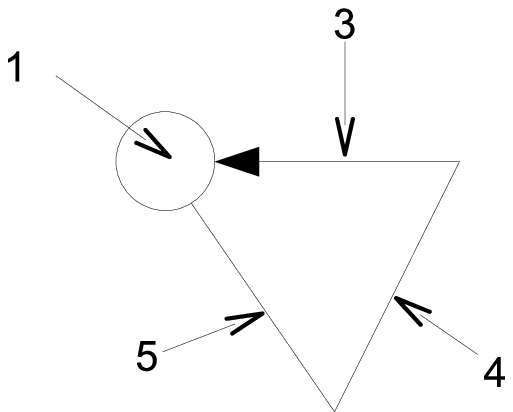}
\\[5mm]
(b)
\end{center}
}
\hfill\strut
\end{center}
\caption{The loop complex (a): the amplifiable fragment (2) and the loop link (1).
The schematic representation (b) highlighting the 3 components of the amplifiable fragment}\label{fig:circular}
\end{figure}

The mechanism is represented on Fig.~\ref{fig:circular}(a). The loop complex is
a circular molecule composed of two parts: the amplifiable fragment (2) and the
loop link (1). The arrow represents the 3' end of the amplifiable fragment. We
represent this molecule schematically as on Fig.~\ref{fig:circular}(b). For the
sake of commodity we split the amplifiable fragment in 3 parts (3,4,5 on the
picture) corresponding to the 3' end, middle and 5' end of the amplifiable
fragment.

\begin{figure}[htb]
\begin{center}
\parbox{30mm}{
\begin{center}
\includegraphics[scale=0.5]{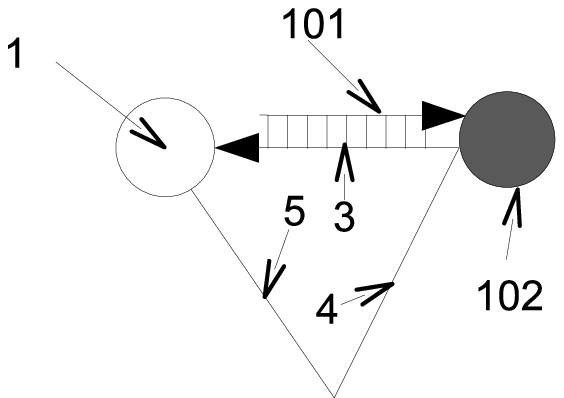}
\\[5mm]
(a)
\end{center}
}
\hfill
\parbox{31mm}{
\begin{center}
\includegraphics[scale=0.5]{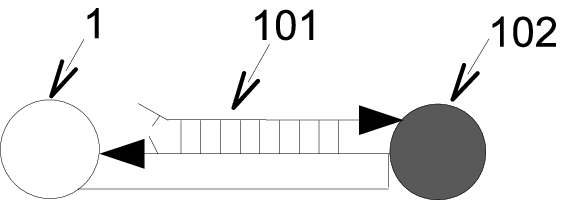}
\\[15mm]
(b)
\end{center}
}
\hfill
\parbox{31mm}{
\begin{center}
\includegraphics[scale=0.5]{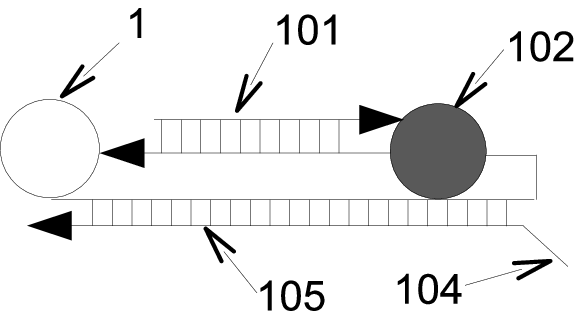}
\\[8mm]
(c)
\end{center}
}
\end{center}
\caption{The amplification process: primer extension (a); maximum stretch of amplifiable fragment
and the opening of
the 5' end of the double strand (b); a second amplification started (c).
Notation: link loop (1), amplifiable fragment (3,4,5), DNA polymerase (102), (extended) primer (101),
single stranded fragment (104).}\label{fig:extension}
\end{figure}

The amplification mix contains primers (101) which hybridize to the 3' part of
the amplifiable fragment (3). They can be further extended by DNA polymerase
(102) present in the mix, see Fig.~\ref{fig:extension}(a). The loop link (1)
length is small compared to the length
of the DNA fragment: typically 1 to 5 nm. It can be a simple chemical link
joining the extremities of the DNA fragment, or a biochemical link between
biotin moieties attached to the extremities of the DNA fragment via a
streptavidin protein. The DNA fragment is also short in regards to the
stiffness of double stranded DNA. In conditions where the biochemical
replication reactions can take place, double stranded DNA molecules shorter
than 
300~--~500 nucleotides are too stiff for their extremities to come into close
proximity. In other words, a circular DNA molecule shorter than 300 nucleotides
cannot exist in full double stranded form, but is found as stretches of double
stranded portions separated by single stranded portions. This is true also for
the loop complexes used in ICLEDA.
 At some point the complex will be composed
from a single stranded DNA having $n$ nucleotides from the 5' end of the
amplifiable fragment,  a double-stranded DNA corresponding to the 3' part of
the amplifiable fragment,  the extended primer and  the linking loop of special
length.


Since the two extremities of the amplifiable fragment are linked to each other
this gives a geometric constraint for the loop. In order to continue the
reaction either the single strand part should be extended to the maximum or the
double stranded part should open at the opposite extremity. At some level of
tension the energetic preference will be to continue the extension of the
primer by DNA polymerase, while the opposite end will detach by Brownian
motion. So, at the same time the double stranded fragment will be opened at 5'
part and one nucleotide will be added by DNA polymerase. However it should be
noted that the number of nucleotides on the double stranded part remains
unchanged, due to the geometric constraints of the loop.

Since no more nucleotides are bound at the 3' end of the amplifiable fragment
(3) at some point it becomes accessible for a hybridization with a new primer,
see Fig.~\ref{fig:extension}(b). The extension is blocked when it reaches the
end on the amplifiable fragment (105) because of the presence of non-natural
nucleotides in the link, see Fig.~\ref{fig:extension}(c).

\begin{figure}[htb]
\begin{center}
\includegraphics[scale=0.7]{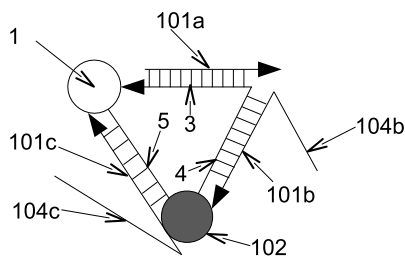}
\end{center}
\caption{The amplification process: three molecules in different stages of duplication.
Notation: link loop (1), amplifiable fragment (3,4,5), DNA polymerase (102), (extended) primers (101a,b,c),
single stranded fragments (104b,c).}\label{fig:extension2}
\end{figure}

Fig.~\ref{fig:extension2} shows a loop complex (1) which has three attached
primers being in different stages of the duplication. The first primer (101a)
is paired to the 3' part of the amplifiable fragment (3) and is ready to be
extended. The second primer (101b) is in the process of the extension and it is
paired by its 3' end to the central part (4) of the amplifiable fragment where
its extension continues, while its 5' part is a single stranded DNA. The third
primer (101c) reached the end and is not extended anymore, but it is still
paired by its 3' end to the 5' end (5) of the amplifiable fragment. The
progression of the extension is blocked by the loop link (1), while its 5' part
is progressively detached from the amplifiable fragment by the progression of
the extension of the second primer. Besides its role in the amplification
process, the link also enables the possibility of attaching the loop complex to
a surface without hindering the replication mechanism.

\begin{figure}[htb]
\begin{center}
\strut\hfill
\parbox{31mm}{
\begin{center}
\includegraphics[scale=0.5]{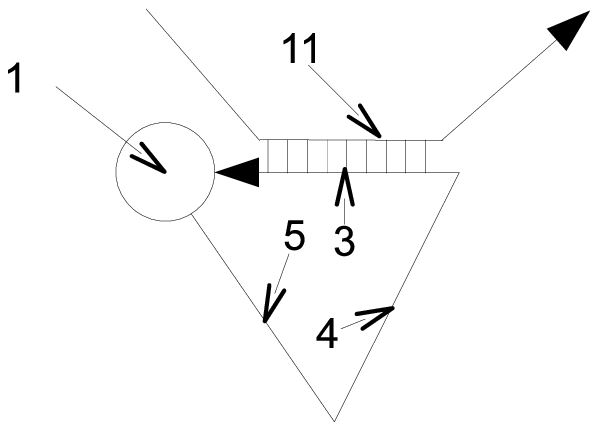}
\\[5mm]
(a)
\end{center}
}
\hfill
\parbox{39mm}{
\begin{center}
\includegraphics[scale=0.5]{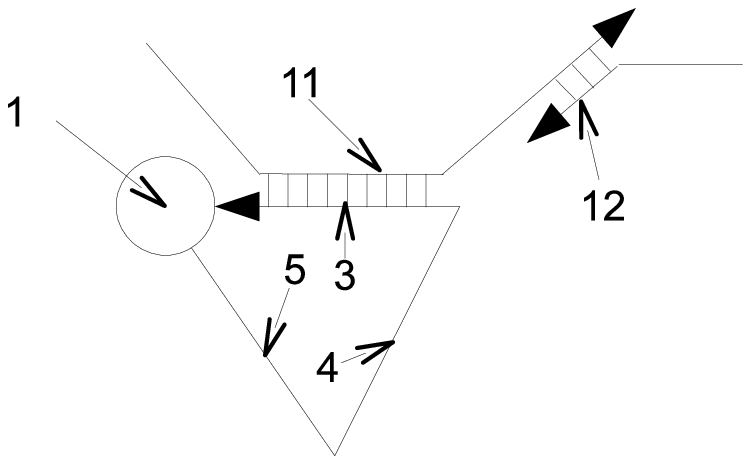}
\\[5mm]
(b)
\end{center}
}
\hfill\strut
\end{center}
\caption{The loop complex blocked by a trigger (a) and the hybridization with further
extension used to remove the trigger (b).
Notation: link loop (1), amplifiable fragment (3,4,5), trigger (11),
DNA strand used to release the trigger (12)}\label{fig:trigger}
\end{figure}

Now we remark that if in the mix a fragment of  a single stranded DNA that
matches the 3' part of the amplifiable fragment is present, then it can stick
to the amplifiable fragment as shown on Fig.~\ref{fig:trigger}(a). We call such
a strand a \emph{trigger}. When a trigger is attached to the loop complex, no
amplification can be done. A trigger can be detached from the loop 
complex 
by an \emph{activator} that matches by its 3' end a part of the trigger strand
as shown on Fig.~\ref{fig:trigger}(b). Once bound to the trigger the activator
can be extended by DNA polymerase and this will release the trigger, so the
loop complex will be able to start the amplification process.

\subsection{Formalization}
\label{ssec:formal}

  As the main process described in the previous subsection deals with
 molecules, we shall represent them as {\it words} over the four-letter alphabet.
However, we shall not represent individual nucleotides. We shall rather
consider the places where reactions may occur. Consequently, we shall divide
the words in several places, the {\bf sensitive} ones and the {\bf neutral}
ones. If $A$~is a molecule or a part of it, we denote by~$A'$ its complement in
the Watson-Crick complementarity rule. Note that~$A'$ is also written in the
opposite order of its letters with respect to~$A$. Molecules are oriented and
we denote by the symbol $\di$ the head (3' end) of the molecule, which implies
the reading order from left to right for $\di A$  and from right to left for
$A'\di$.

    We consider that if $A$ and~$A'$ are both present and if the
configurations of the DNA strings to which they belong allow it, they bind each
other. Consider that  $A$ occurs in a molecule~$M$. We write this $M=uAv$
with~$u$ or~$v$ possibly not present: we then say that $u$ or~$v$ is empty.
Consider that $A'$ occurs in a molecule~$M'$. We similarly denote this by
$M'=xA'y$. We assume that $A$~no more 
occurs neither in~$u$ nor in~$v$ and
that, similarly $A'$~no more occurs neither in~$x$ nor in~$y$. We also assume
that $A$ and~$A'$ do not interact with neither of the molecules $u$, $v$, $x$
and~$y$ and that these molecules also do no interact with each other. We
express this by saying that $u$, $v$, $x$ and~$y$ are {\bf neutral} parts of
the molecules to which they belong. This allows us to focus only on~$A$
and~$A'$ which are called the {\bf sensitive} parts of the molecules~$M$
and~$M'$. We shall mark this difference between sensitive and neutral parts of
molecules in the notation: sensitive parts will be denoted by capital letters
and neutral parts will be denoted by lower case ones.

    When the molecules $\diamond M$ and~$N\diamond$
are both present, assuming that $\diamond M$ contains the active part~$A$ and
that $N\diamond$ contains~$A'$, we denote this by an additive notation:
\hbox{$\diamond M~\bigoplus~N\diamond$}. Now, if we replace $\diamond M$
and~
$N\diamond$ by their expressions in terms of~$A$ and~$A'$, we get
\hbox{$\diamond uAv~\bigoplus~yA'x\diamond$} and we now know that as a result
we obtain a complex as $A$ and~$A'$ get bound to each other. We write the
complex as \complex{\diamond uAv}{yA'x\di}. Hence the corresponding rule can be
written as follows:

\begin{equation}\label{eq:complex}
\di uAv\op yA'x\di\vdash\complex{\di uAv}{yA'x\di}
\end{equation}

As an example, we cannot write
\begin{equation}\label{eq:badcomplex}
\di uAv\op yBx\di\vdash\complex{\di uAv}{yBx\di}
\end{equation}
unless $A=wB't$ or $B=rA's$, we remember the reader that lower case letters
denote neutral parts.
To avoid unneeded repetition of rules we shall always use the rule
in its explicit form~\eqref{eq:complex},considering that in~\eqref{eq:badcomplex}, we have
neither $A=wB't$, nor~$B=rA's$.

We assume that the operation $\op$ is commutative and associative,
which corresponds to the fact that $\op$~models a situation
in which the components are independent and may freely combine or not
and in all possible combinations.

The loop complex can be formalized as follows: \loopc{F'uR'} where $u$ is the
neutral part and $F'$ with~$R'$ are the sensitive ones, corresponding to the
parts 4,3,5 on Fig.~\ref{fig:circular}(b). A trigger can be formalized as $\di
wX'Fz$, where $w$ and~$z$ are the neutral parts and $X'$, $F$ are the sensitive
ones.

The working of the loop complex can be formalized as:

\begin{equation}\label{eq:loop}
\loopc{F'uR'}\To Fu'R.
\end{equation}

Now, if there is a trigger, we have:

\begin{equation}\label{eq:close loop}
\loopc{F'uR'}\op \di tA'Fw \To \complex{\loopc{F'uR'}}{\di tA'Fw}.
\end{equation}

\noindent
where $u$, $t$ and~$w$ are neutral. From subsection~\ref{ssec:ampli_loop},
the result of~\eqref{eq:close loop} blocks the application of~\eqref{eq:loop}.

We also have two rules for the trigger which occurs in formula~\eqref{eq:close loop}:

\begin{align}\label{eq:open loop1}
\complex{\loopc{F'uR'}}{\di tA'Fw}\op zF'Ax\di &\To \loopc{F'uR'}\op \complex{\di tA'Fw}{zF'Ax\di}\\
\label{eq:open loop2}
\complex{\loopc{F'uR'}}{\di tA'Fw}\op Ax\di &\To \loopc{F'uR'}\op \complex{\di tA'Fw}{F'Ax\di}
\end{align}

\noindent In these formulas, as $A$ binds with~$A'$ which, as a result,
detaches the trigger from the loop complex, see Fig.~\ref{fig:trigger}. And
now, we can see that (3)~applies. Note that both formulas~\eqref{eq:open loop1}
and~\eqref{eq:open loop2} produce almost the same result as the molecule
$F'Ax\diamond$ is present in the new complex in both cases. The
formula~\eqref{eq:open loop2} translates the property indicated in
Sub-section~\ref{ssec:ampli_loop}: to detach the trigger attached to the
complex in the left-hand side of the formulas, it is enough to present the
beginning of the active molecules, the process of detachment will produce the
continuation of the active parts.

   It can be noted that the formalism allows to explain why we obtain this rule.
Indeed, it might be argued that as $F'$ and~$F$ should also bind together,
we could have the backward reaction:
\begin{equation*}
\loopc{F'uR'}\op \complex{\di tA'Fw}{zF'Ax\di}\To \complex{\loopc{F'uR'}}{\di tA'Fw}\op zF'Ax\di
\end{equation*}
The reason why we have not this reaction is that as $A'$ is more visible
for~$A$ that $F$ is for~$F'$, $A$~attaches to the trigger which, as a result,
lead to its freeing from the complex.

   In what follows, to simplify the notations, we introduce the following
convention: we represent the triggers $Ax\diamond$ and $zF'Ax\diamond$
as~$\si{A}$. This allows us to melt the formulas \eqref{eq:open loop1} and 
\eqref{eq:open loop2} in a single
one,namely:

\begin{equation}\label{eq:open loop A}
\complex{\loopc{F'uR'}}{\di tA'Fw}\op \si{A} \To \loopc{F'uR'}\op \complex{\di tA'Fw}{\si{A}}
\end{equation}

\noindent
If we have to explicit the form of~$\si{A}$ as $Ax\diamond$ or $zF'Ax\diamond$,
we shall speak of a {\bf realization} of~$\si{A}$.

We can see this higher priority of~$\si{A}$ in the formalism. Remember that
if we wish to read the words in the order given by a run other the molecule
from its head to its tail, we have to read the word from the diamond to
the opposite end.  Define the {\bf apartness} of a
sensitive molecule~$X$ with respect to a molecule $M$ as
the number of sensitive parts of~$M$ between the diamond of~$M$ and
the position of~$X$. Denote it 
by $apart(X,M)$. We can now define
the {\bf apartness} of~$X$ with respect to~$M$ and~$N$, denoted by
$apart(X,M,N)$, where $X$~is
contained in~$M$ and $X'$ is contained in~$N$, as the expression
\hbox{$apart(X,M)+apart(X',N)$}. Now,
we can see that
\hbox{$apart(\si{A},\complex{\loopc{F'uR'}}{\di tA'Fw},\si{A})=0$}
whatever the realization of~$\si{A}$: as $Ax\diamond$ or $zF'Ax\diamond$.
Now we have that
\hbox{$apart(F,\complex{\loopc{F'uR'}}{\di tA'Fw},\si{A})=1$},
also whatever the realization of~$\si{A}$.
As the apartness of~$\si{A}$ is lower than
that of~$F$, the reaction with~$\si{A}$ has a higher priority and so it takes place
while the reaction with~$F$ does not.

\section{Implementing boolean functions}

In this section we consider the amplification loop $\loopc{F'uR'}$.  We assume
that if a loop $\loopc{F'uR'}$ is unblocked, then there will be an
\textbf{unbounded} number of copies of $Fu'R$. This assumption results from the
observation that once started, the amplification could produce a large enough
number of resulting molecules, even if the loop is blocked again afterwards.

This implies that we can consider that initially all loops are blocked,
otherwise we substitute them by a large number of DNA molecules corresponding
to their result. So the computation in such a system consists in unblocking
some loops in some order. This corresponds in a direct manner to boolean
circuits where the electrical impulses are propagated in the circuit. The
signals we use are always identified by the part at the beginning of the
molecule, \ie{} a signal $\si{A}$ will be given by the string $\di Aw$.

It is known that any boolean function can be computed by a boolean circuit with
AND, OR and NOT gates. It is possible to omit the NOT gate if signals
corresponding to false values of variables are used. We remark that in this
case only one type of signal (corresponding to \textbf{true}) exists in the
circuit and the result is obtained by positional information, \ie{} there are 2
possible output wires corresponding to the true and the false values of the
formula.

   From the properties stated in the previous section, we can devise
the construction of the $OR$- and the $AND$-gates. First, we start with the
$OR$-gate described in Subsection~\ref{ssec:orgate}. In
Subsection~\ref{ssec:andgate1}, we describe an $AND$ gate which is close to the
construction of Subsection~\ref{ssec:orgate}. In
Subsection~\ref{ssec:andgate2}, we describe another variant which is closer to
that suggested by~\cite{brevet}.

\subsection{The $OR$-gate}
\label{ssec:orgate}

   Considering rule~\eqref{eq:loop}, we decide to interpret the production of
the molecule~$\si{R}$ as the emission of the boolean signal {\tt true}.

   In this condition, if the trigger \hbox{$tA'Fw\diamond$} is initially attached to
a loop complex \loopc{F'uR'}, forming 
complex
\complex{\loopc{F'uR'}}{\diamond tA'Fw}, 
rule~\eqref{eq:open loop A} tells us that introducing the molecule~$\si{A}$
either as $Ax\diamond$ or as $zF'Ax\diamond$,
we obtain
\hbox{$\loopc{F'uR'}\op \complex{\di tA'Fw}{\si{A}}$}.
As $\bigoplus$ is commutative, this corresponds to the fact that now
rule~\eqref{eq:loop} applies to \loopc{F'uR'}: accordingly, we get again the
signal {\tt true}.

\begin{figure}[htb]
\begin{center}
\includegraphics[scale=0.5]{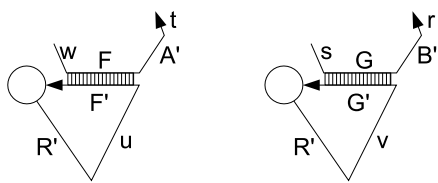}
\end{center}
\caption{The simulation of the OR gate.}\label{fig:or}
\end{figure}

It is not difficult to obtain a similar sequence of deductions with
\loopc{G'vR'} and the trigger \hbox{$\diamond rB'Gs$}. Introducing the
molecule~$\si{B}$, either as~$By\diamond$ or as $zG'Bx\diamond$, we shall also
get~$\si{R}$ by applying the rules. The final construction for the gate is
shown on Fig.~\ref{fig:or}.

It is plain that if we introduce one of~$\si{A}$ or~$\si{B}$ we get
also~$\si{R}$: it is enough to introduce initially in the soup with both the
loop complexes \loopc{F'uR'} and \loopc{G'vR'} as well as both triggers
\hbox{$\diamond tA'Fw$} and \hbox{$\diamond rB'Gs$}, so that we may assume
that, after a certain time, rule~\eqref{eq:close loop} applies giving us:
\begin{equation*}
\complex{\loopc{F'uR'}}{\di tA'Fw}\op \complex{\loopc{G'vR'}}{\di rB'Gs}.
\end{equation*}

>From rule~\eqref{eq:open loop A} and the commutativity and associativity, we
obtain that if $\si{A}$ is poured into the soup, $\si{R}$ will be produced, as
the complex \complex{\loopc{F'uR'}}{\di tA'Fw} will be decomposed into
\hbox{$\loopc{F'uR'}\op\complex{\di tA'Fw}{\si A}$}.
If ~$\si{B}$ is poured, we get a similar result.
If both $\si{A}$ and~$\si{B}$ are introduced, we get a similar result as
both kinds of complexes are present. In all these three cases,
the molecule~$\si{R}$ is produced.

Accordingly, we obtain that a soup which contains both the loop complexes
\complex{\loopc{F'uR'}}{\di tA'Fw} and \complex{\loopc{G'vR'}}{\di rB'Gs},
it behaves like an $OR$-gate with respect to the occurrence
or absence of the molecules~$\si{A}$ and~$\si{B}$: the required signal~$\si{R}$ occurs if at
least one of them is present and only in this condition.

We remark that the construction above can be extended to an $n$-ary OR gate.
This gives the possibility to simulate semi-unbounded fan-in circuits.

\subsection{The $AND$-gate}
\label{ssec:andgate1}

We can simulate an AND gate by considering two active regions on the loop,
\ie{} loops of form $\loopc{F_A'uF_B'vR'}$, where $F_A'$ and $F_B'$ are active
zones for triggers having $A$ and $B$. Then the loop is blocked by two
molecules as follows \complex{\complex{\loopc{F_A'uF_B'vR'}}{\di tA'F_Aw}}{\di
qB'F_Bs}, see Fig.~\ref{fig:and}(a). Now if both triggers $\di At'$ and $\di
Bq'$ are present, then the loop complex can be unblocked:

\begin{multline}\label{eq:and1}
\complex{\complex{\loopc{F_A'uF_B'vR'}}{\di tA'F_Aw}}{\di qB'F_Bs}\op
\di At'\op \di Bq'\To\\
\To \complex{\loopc{F_A'uF_B'vR'}}{\di tA'F_Aw}\op \di At'\op
\complex{\di qB'F_Bs}{\di Bq'}\To \\
\To \loopc{F_A'uF_B'vR'}\op \complex{\di tA'F_Aw}{\di At'}\op \complex{\di
qB'F_Bs}{\di Bq'}
\end{multline}

In the 
above 
derivation 
the unblocking can start by the signal $At'$, but
following the commutativity of $\op$ it yields the same result.

It is clear that if only one of the triggers $At'$ or $Bq'$ is present, then
the loop complex is only partially unblocked (either
$\complex{\loopc{F_A'uF_B'vR'}}{\di tA'F_Aw}$ or
$\complex{\loopc{F_A'uF_B'vR'}}{\di qB'F_Bs}$) and cannot produce the resulting
signal.

\subsection{The initial $AND$-gate}
\label{ssec:andgate2}

\begin{figure}[htb]
\begin{center}
\strut\hfill
\parbox{30mm}{
\begin{center}
\includegraphics[scale=0.5]{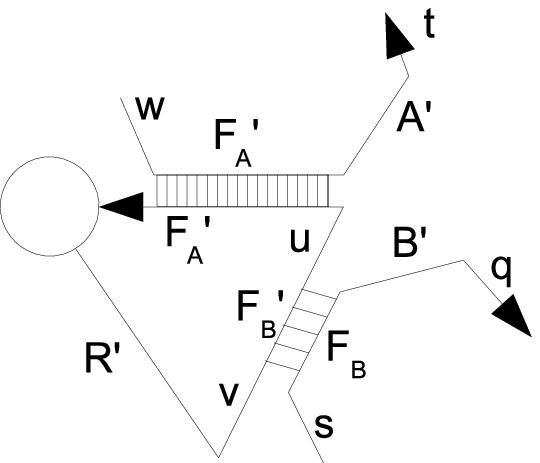}
\\[5mm]
(a)
\end{center}
}
\hfill
\parbox{30mm}{
\begin{center}
\includegraphics[scale=0.5]{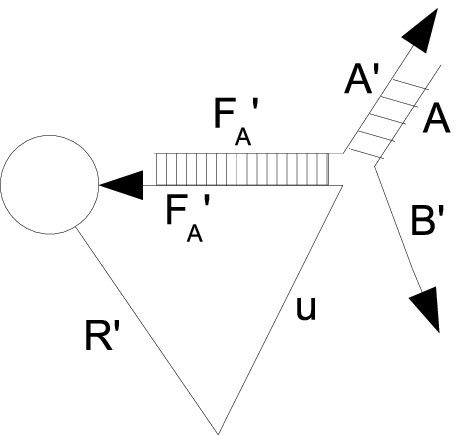}
\\[5mm]
(b)
\end{center}
}
\hfill\strut
\end{center}
\caption{The simulation of the AND gate with two active regions on the loop (a) and
with one active region on the loop (b)}\label{fig:and}
\end{figure}

Another variant of the AND gate is described in~\cite{brevet}. Like in the
previous case it is also a complex of 3 molecules, however the loop complex is
bound in only one place. Its construction is done in two stages: the loop
complex \loopc{F_A'wR'} is blocked by the trigger $\di F_AA'$. After that the
molecule $AB'\di$ is added into 
the 
solution and it will stick to the $A'$ site.
Hence, the complex \complex{\complex{\loopc{F_A'wR'}}{\di A'F_A}}{AB'\di} will
be formed, see Fig.~\ref{fig:and}(b). We remark that since this complex is
formed during the preparation stage, we can insure that no molecules $\di Au$,
($u\ne B$) or $\di tF_AAv$ are present in the solution.

Now during the computation the loop complex can be unblocked as follows:

\begin{multline}\label{eq:and2}
\complex{\complex{\loopc{F_A'wR'}}{\di A'F_A}}{AB'\di} \op \di Bu \op \di
Av
\To \\
\To \complex{\loopc{F_A'wR'}}{\di A'F_A} \op \complex{AB'\di}{\di A'Bu}
\op \di Av\To\\
\To \loopc{F_A'wR'}\op\complex{AB'\di}{\di A'Bu}\op \complex{\di
A'F_A}{\di F_A'Av}
\end{multline}

We remark that unlike the previous case this construction is not symmetric,
\ie{} first the signal $\di Bu$ is removing the molecule $AB'\di$ from the
complex, freeing the site $A'$, which can be bound after that by the signal
$\di Au$ that finally unblocks the loop.

\subsection{Implementing boolean functions}

It is known that every boolean function of $n$ variables can be implemented by
a boolean circuit using AND, OR and NOT gates. It is possible to eliminate the
NOT gate by considering that only true signals can circulate in the circuit. In
this case the input of the circuit is not the true or false value for 
the 
same
variable $x$, but rather a true value for $x$ or for $\neg x$. The output is
also modified: instead of a single output having one of the values true or
false, there are two outputs (marked by true and false) and a true value in
some of the outputs indicate that the output value of the circuit is true or
false.

For a boolean formula $\phi$ such a modified circuit can be constructed by a
superposition of two circuits, one computing $\phi$ and the other one computing
$\neg\phi$.

If we consider that the two output nodes are combined into a fictive output
node then such a circuit is a DAG with the root being the output node and the
leaves being the input variables and their negations.

We remark an important similarity between traditional electronic implementation
of boolean circuits and our implementation: if a signal (represented by an
electric charge in electronics and by DNA molecule in our case) appears at some
moment during the computation, then it is sufficiently strong and does not
disappear in the consequent steps. This 
allows us to make a direct analogy
between two implementations and use similar construction techniques. This is
different from other approaches of simulation of circuits by DNA computing, as
we do not need additional amplification phase anymore.

More precisely, let $f:B^n\to B$, $B=\{0,1\}$ be a boolean function and let
$D=(V,E,F)$ be the circuit implementing this function (where $V=\{1,\dots,n\}$
is the set of vertices, $E\subseteq V\times V$ is the set of edges and $F:V\to
\{x_p,\neg x_p,AND_m,OR_k,\linebreak NOT_q,out\}$, $1\le p,q,m,k\le n$ is the
function that labels vertices of the circuit). Then for any inner node $y$ such
that $F(y)=OR_m$ (resp. $F(y)=AND_m$) we construct an OR loop (resp. AND loop)
as discussed in~\ref{ssec:orgate} (resp.~\ref{ssec:andgate1}). The final gate
will send the signal to the output node (the root of the circuit ($F(x)=out$)
). A similar construction should be performed for $\neg f$. The final assembly
is the union of these two circuits.

In order to compute the result in the initial configuration signals
corresponding to $X_k$ (where $X_k$ is either $x_k$ or $\neg x_k$) should be
introduced.

We give below an example of such a construction for  the following function:
$f(x_1,x_2,x_3)=(\neg x_1\wedge x_2) \ \vee\ (x_1\wedge x_2\wedge\neg x_3)$.

The ordinary boolean circuit computing $f$ is given below:

$$
\xymatrix@=3mm{
x_1 \ar@{-}[dddrr]\ar@{-}[rr]&&NOT\ar@{-}[d]\\
  &  & AND\ar@{-}[dr]\\
x_2 \ar@{-}[urr]\ar@{-}[ddr]  &&& OR\ar@{-}[dr] \\
&& AND\ar@{-}[ur] && out\\
x_3\ar@{-}[d]&  AND\ar@{-}[ur]&&\\
NOT \ar@{-}[ur]&\\
}
$$

Next we replace the NOT gate by considering that only true values can transit
the circuit. This gives the following structure for $f$ (we also numbered the
gates):

$$
\xymatrix@=3mm{
x_1 \ar@{-}[dddrr]&\\
\neg x_1 \ar@{-}[rr]  &  & AND_1\ar@{-}[dr]\\
x_2 \ar@{-}[urr]\ar@{-}[ddr]  &&& OR_4\ar@{-}[dr] \\
&& AND_3\ar@{-}[ur] && out\\
&  AND_2\ar@{-}[ur]&&\\
\neg x_3 \ar@{-}[ur]&\\
}
$$

Now we should construct the circuit (without negation) for $\neg f=(x_1\wedge
x_3)\vee \neg x_2$:

$$
\xymatrix@=3mm{
x_1 \ar@{-}[ddrr]&\\
 &  &\\
  &&AND_5\ar@{-}[ddr]&  \\
\neg x_2\ar@{-}[drrr]&&  && out\\
x_3\ar@{-}[uurr]&  &&OR_6\ar@{-}[ur]\\
&\\
}
$$

Now we combine the two constructions. We also add labels to edges going out
from the same left node (corresponding to a concrete signal produced by the
corresponding gate).

$$
\xymatrix@=5mm{
x_1 \ar@{-}[dddrr]\ar@{-}[ddrr]^<<<<<<<<{S_1}&\\
\neg x_1 \ar@{-}[rr]_<<<<{S_2}  &  & AND_1\ar@{-}[dr]^{S_7}\\
x_2 \ar@{-}[urr]_<{S_3}\ar@{-}[ddr]  &&AND_5\ar@{-}[ddr]^>>>>>>>{S_8}& OR_4\ar@{-}[r]^{S_{11}} & out_{true}\\
\neg x_2\ar@{-}[drrr]_<<<{S_4}&& AND_3\ar@{-}[ur]^>>>>{S_{10}} && out_{false}\\
x_3\ar@{-}[uurr]^{S_5}&  AND_2\ar@{-}[ur]_<<<{S_9}&&OR_6\ar@{-}[ur]_{S_{12}}\\
\neg x_3 \ar@{-}[ur]^{S_6}&\\
}
$$

Having in mind that the true value of some node is represented by the presence
of the corresponding signal (that labels the edge), it becomes clear that this
construction can be directly implemented using loop complexes and corresponding
signals by 4 AND gates and 2 OR gates.

The signals $S_1-S_6$ correspond to the input values and the signals $S_{11}$
and $S_{12}$ to the output. So the computation starts by giving input signals
(taking care of not having an input $x$ and $\neg x$ at the same time). Then
the gates will act in cascade and one of two output signals ($S_{11}$ if $f$ is
true or $S_{12}$ if $f$ is false) will be obtained.


\section{Conclusions}

In this article we present a new method for the simulation of boolean circuits.
The use of ICLEDA offers 
many advantages like a single volume and unchanged
reaction conditions. This implies that the corresponding implementation will
not need any additional intervention. Moreover, since the loop complexes can be
easily attached to the support it is possible to reuse the circuit by washing
the tube and by introducing trigger molecules to block the loops. Another
advantage of the method is that the signal molecules (corresponding to the true
value of some gate) are of a small length; moreover, by introducing
compartments it is possible to share some of the signals.

As a further work it remains to experimentally verify the functioning of the
method. A partial attempt for this is done in~\cite{brevet} where an evidence
of the feasibility of the basic blocs is given. Another interesting question is
the further investigation of the framework for the partial hybridization of DNA
strings introduced in this article.

\end{document}